\documentstyle[editedvolume, psfig]{crckapb} 

\begin{opening}
\title{Non-detection of kHz QPOs in GX 9$+$1 and GX 9$+$9}\\
\author{Rudy Wijnands}
\author{Michiel van der Klis}
\institute{Astronomical Institute, University of Amsterdam}
\author{Jan van Paradijs}
\institute{Astronomical Institute, University of Amsterdam, and\\
Department of Physics, University of Alabama at Huntsville}
\end{opening}

\begin{document}

\section{Introduction}
In numerous low-mass X-ray binaries quasi-periodic oscillations (QPOs)
between 300 and 1200 Hz have been discovered (the kHz QPOs; see van
der Klis 1997 for a recent review on kHz QPOs). Here we present the
search for kHz QPOs in the atoll sources GX 9$+$1 and GX 9$+$9.

\section{Observations and analysis}

We observed GX 9$+$1 on 1996 Feb 29, Apr 21, May 29, and 1997 Feb 10
and Mar 9, and GX 9$+$9 on 1996 Aug 12, Oct 16, and Oct 30 with the
RXTE satellite. We obtained a total of 23.3 ksec (GX 9$+$1) and 15.2
ksec (GX\,9$+$9) of data.  The X-ray hardness-intensity diagrams
(HIDs) were made using the {\it Standard 2} data.  Due to gain changes
the GX 9$+$1 HID for the 1996 Feb 29 observation can not be directly
compared with those of the other observations.  The power density
spectra were made using the 250$\mu$s time resolution data. We
calculated rms amplitude upper limits (95\% confidence) on QPOs with a
FWHM of 150 Hz in the frequency range 100--1500 Hz.

\section{Results}

The HIDs for GX 9$+$1 and GX 9$+$9 are shown in Figure 1. According to
the HID (Fig. 1a) and the high-frequency noise in the power spectrum
GX 9$+$1 was on the lower banana during the 1996 Feb 29
observation. During the other observations GX 9$+$1 moved along the
banana branch (Fig. 1b). The power spectrum and the HID of GX\,9$+$9
suggest that this source was on the banana branch during the
observations.  We find for GX 9$+$1 upper limits of 1.6\% (the 1996
Feb 29 observation; energy range 2--60 keV) and 1.3\% (all other
observations combined; energy range 2--18.2 keV), and for GX 9$+$9 an
upper limit of 1.8\% (all data; energy range 2--18.2 keV). We divided
the GX 9$+$1 banana in Figure 1b into different regions. For each
region we calculated the rms upper limit on kHz QPOs. We find rms
amplitude upper limits of 3.2\%, 1.3\%, 1.9\%, 2.7\%, and 3.4\%
(energy range 2--18.2 keV), for region 1, 2, 3, 4, and 5,
respectively.

\begin{figure}
\psfig{figure=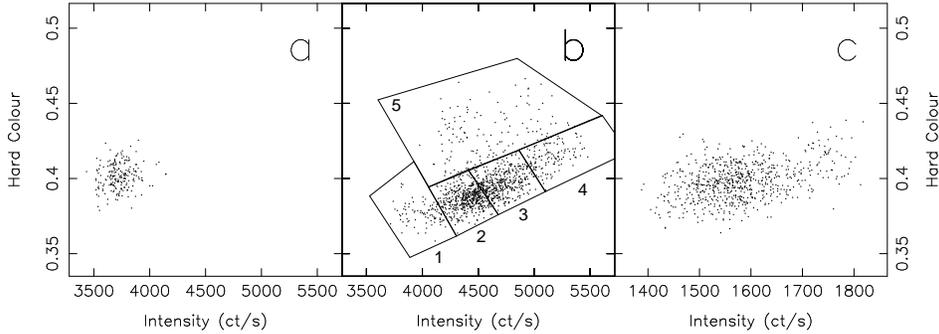,width=12.5cm}
\caption{The HIDs of GX 9$+$1 ({\it a} and {\it b}) and GX 9$+$9 ({\it
c}). The data of 1996 Feb 29 of GX 9$+$1 ({\it a}) were taken with a
different PCA gain compared to the data of the other observations
({\it b} and {\it c}). The intensity is the count rate in the photon
energy range 2.0--15.9 keV ({\it a}) or 2.1--16.0 keV ({\it b} and
{\it c}); the hard colour is the count rate ratio between 9.7--15.9 kev
and 6.5--9.7 keV in {\it a}, and between 9.7--16.0 keV and 6.4--9.7
keV in {\it b} and {\it c}.  All points are 16s averages. The count
rates are background subtracted, but not dead-time corrected. The
regions in {\it b} have been used to calculate the upper limits.}
\end{figure}

\section{Discussion}

The non-detection of kHz QPOs in GX 9$+$1 and GX 9$+$9 is consistent
with the predictions of the sonic-point model proposed to explain the
kHz QPOs (Miller et al. 1997). It is known from other atoll sources
(e.g. 4U 1636$-$53: Wijnands et al. 1997; 4U 1820-30: Smale et
al. 1997) that when they are in the upper banana branch the kHz QPOs
are not detected.  Thus, it remains possible that when GX 9$+$1 and GX
9$+$9 are observed longer on the lower banana, or even in the island
state, kHz QPOs are detected in these sources.

\end{document}